\documentclass[12pt,preprint]{aastex}
\shortauthors{\"Oberg et al.}

\begin{document}

\title{Evidence for Multiple Pathways to Deuterium Enhancements 
in Protoplanetary Disks}

\author{Karin I. \"Oberg\altaffilmark{1}, Chunhua Qi, David J. Wilner}
\affil{Harvard-Smithsonian Center for Astrophysics, 60 Garden Street, 
Cambridge, MA 02138, USA}
\email{koberg@cfa.harvard.edu}
\author{Michiel R. Hogerheijde}
\affil{Leiden Observatory, Leiden University, Post Office Box 9513, 2300 RA Leiden, Netherlands}

\altaffiltext{1}{Hubble Fellow}

\begin{abstract}

The distributions of deuterated molecules in protoplanetary disks are 
expected to depend on the molecular formation pathways. 
We use observations of spatially resolved DCN emission from the 
disk around TW Hya, acquired during ALMA Science verification
with a $\sim3''$ synthesized beam, together with comparable DCO$^+$ 
observations from the Submillimeter Array, to investigate differences 
in the radial distributions of these species and hence differences in their formation 
chemistry. In contrast to DCO$^+$, which shows an increasing 
column density with radius, DCN is better fit by a model that is 
centrally peaked.  We infer that DCN forms at a smaller radii and 
thus at higher temperatures than DCO$^+$. This is consistent with 
chemical network model predictions of DCO$^+$ formation from H$_2$D$^+$ 
at T$<$30~K and DCN formation from additional pathways involving 
CH$_2$D$^+$ at higher temperatures.
We estimate a DCN/HCN abundance ratio of $\sim0.017$, similar to the
DCO$^+$/HCO$^+$ abundance ratio. Deuterium fractionation appears to be 
efficient at a range of temperatures in this protoplanetary disk. These
results suggest caution in interpreting the range of deuterium fractions 
observed in Solar System bodies, as multiple formation pathways should be 
taken into account. 
 
\end{abstract}


\section{Introduction}

Deuterium fractionation in interstellar cloud cores, protostars and 
Solar System bodies is frequently used to infer important aspects of their 
physical and chemical histories \citep{Brown88,Herbst09,Mumma11}. 
The deuterium enhancement in the Earth's sea water, with respect to the cosmic abundance,
has been used as an argument for volatile delivery to the young Earth from cold, comet-like planetesimals, even though most present-day comets have about a factor two higher D/H ratio compared to the Earth's sea water \citep{Mumma11}. Recent observations of Earth-like deuterium fractionation 
in a Jupiter family comet \citep{Hartogh11} has lent further support to this scenario. 
Deuterated molecules, notably DCO$^+$ and DCN, are also detected in a handful 
of protoplanetary disks, the sites of ongoing planet formation, 
with abundances that imply deuterium enhancements by orders of magnitude 
\citep{vanDishoeck03,Guilloteau06,Qi08,Oberg10c,Oberg11d}. Understanding how and under 
what conditions these deuterated molecules form in disks is key to 
use deuterium fractionation to probe physical conditions, and to explain 
the deuterium fractionation pattern in 
the Solar System. The formation chemistry of DCN is of particular 
interest since the DCN/HCN ratio can be measured in both comets \citep[e.g.][]{Meier98} and disks, enabling direct comparison.

Deuterium levels in molecules are enhanced above the cosmic deuterium 
abundance of $\sim2\times10^{-5}$ because of differences in zero-point energy 
between deuterated species and the non-deuterated equivalents. 
The reverse reactions are efficiently inhibited at low temperatures, 
resulting in deuterium fractionation.
However, the ``low temperature'' regime is different for 
different reactants. Deuterated species that form from reactions with 
H$_2$D$^+$, in particular DCO$^+$, are expected to be abundant only at
temperatures below 30~K \citep{Willacy07} because of the modest activation 
energy of 230~K for converting H$_2$D$^+$ back into H$_3^+$ \citep{Gerlich02}. 
The activation energy barrier for the conversion between CH$_2$D$^+$ and 
CH$_3^+$ is higher, 390~K \citep{Asvany04}, and CH$_2$D$^+$ can therefore drive a warmer 
deuterium chemistry. Theoretically, DCN may form through reactions with the 
CH$_2$D$^+$ reaction product CHD, as well as with H$_2$D$^+$, DCO$^+$,  D, 
and through grain surface reactions followed by thermal or non-thermal 
desorption \citep{Aikawa01,Willacy07}. Observations of warm DCN in the Orion bar provides observational evidence for that the CH$_2$D$^+$ reaction pathway is important under some interstellar conditions \citep{Parise09}. The formation pathways of DCN that 
are efficient in disks have yet to be constrained by observation. 

TW Hydrae is the most well-studied protoplanetary disk because of its 
proximity (51~pc) and near face-on viewing geometry that allows 
for direct investigation of the radial distribution of molecular emission. 
Millimeter emission from CO, HCO$^+$ 3-2, DCO$^+$ 3-2, HCN 3-2 and 
DCN 3-2 lines have been spatially resolved using the Submillimeter Array 
(SMA). Analysis of DCO$^+$ emission showed that the DCO$^+$ column 
density {\em increases} with radial distance from the star, out to 
$\sim90$~AU, consistent with the low-temperature formation pathway 
from H$_2$D$^+$ \citep{Qi08}. The DCN emission was too weak for detailed 
modeling, but newly released ALMA science verification data on DCN 3-2 
towards TW Hydrae have higher signal-to-noise and allow for significant 
constraints.
% on the DCN radial column density dependence.
In this paper, we use these new data to 
compare the DCN and DCO$^+$ distributions, and to test if these data  
suggest different pathways to deuterium enhancements in these species 
as predicted by some models of disk chemistry. 
%\citep{Aikawa01,Willacy07}.

\section{Spatially Resolved Observations of DCN and DCO$^+$ towards TW Hya}

TW Hya was observed on April 20th and 23rd, 2011 in ALMA band 6 as a part of 
ALMA Science Verification. The calibrated visibilities and CLEANed 
reference images were released and made available to the public in 
August 2011 
\footnote{https://almascience.nrao.edu/almadata/sciver/TWHyaBand6/}. 
These observations include the DCN J=3--2 line at 217.238~GHz, which 
was acquired with a channel spacing of 0.16~km~s$^{-1}$ (resampled to 
0.2~km~s$^{-1}$) and synthesized beam size $2\farcs8\times2\farcs3$,
with an rms of 12~mJy~beam$^{-1}$ in the line-free channels. 
These data are comparable in resolution to the previous observations of 
\citet{Qi08}, with a noise level 14 times better for the ALMA data compared to the SMA observations.

\begin{figure}[htp]
\centering
\epsscale{1}
\plotone{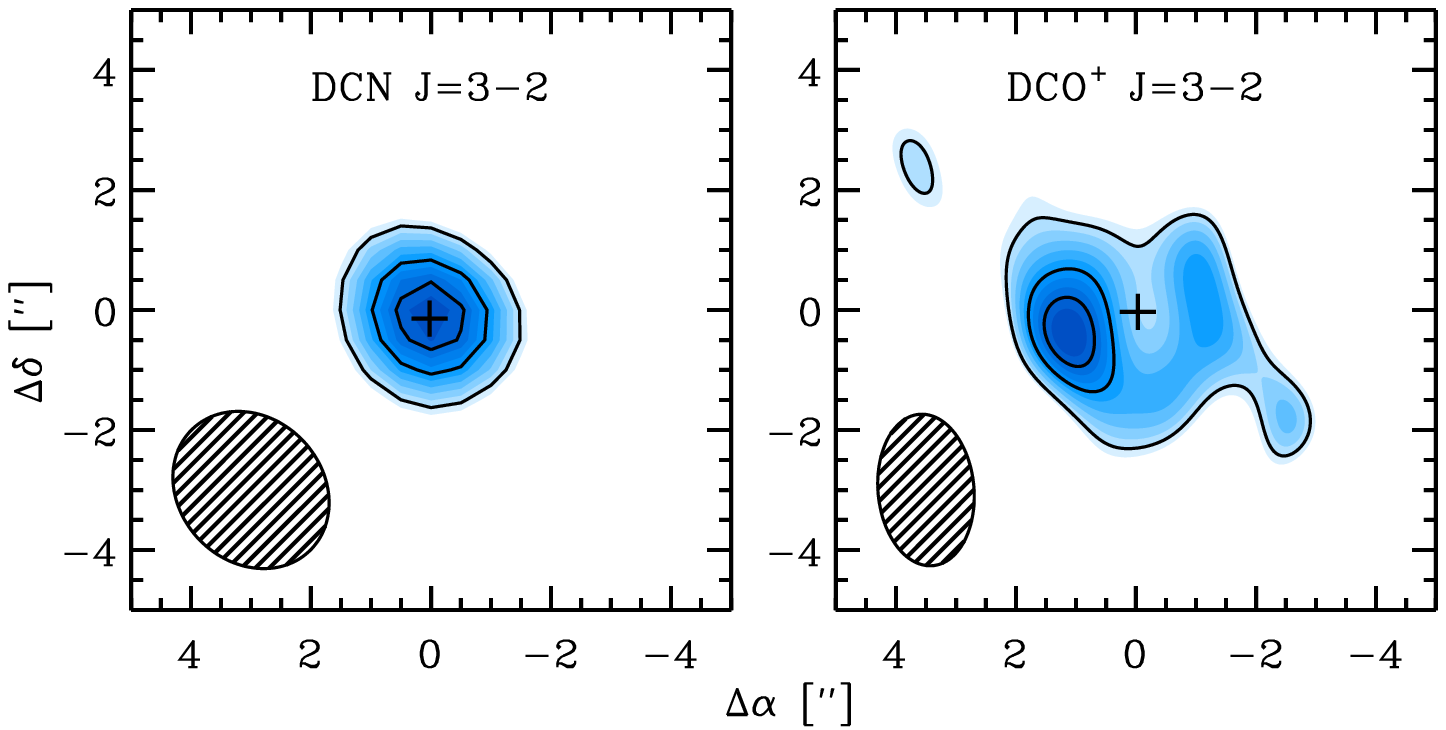}
\caption{
Integrated images of the TW Hya disk in 
({\em left:}) DCN J=3-2 emission (ALMA science verification)
and ({\em right:}) DCO$^+$ J=3-2 emission \citep[SMA data from][]{Qi08}.
The contour levels are at 50\%, 75\% and 90\% of the peak value.
The cross marks the peak of the continuum emission, which locates 
the position of the central star.
\label{fig1}}
\end{figure}

Figure~\ref{fig1} displays the integrated DCN 3-2 emission toward TW Hya. 
The DCN emission is peaked on the TW Hya stellar position and spatially 
compact, barely resolved with the ALMA beam. For comparison, 
Figure~\ref{fig1} shows the integrated DCO$^+$ 3-2 emission, which does 
not peak at the stellar position and appears more extended. Using the DCO$^+$ velocity channel maps, especially the peanut-shaped central one that provides strict constraints on the DCO$^+$ radial distribution, \citet{Qi08} demonstrated that the DCO$^+$ column density increases with distance from the central star out to 90~AU.
Since the excitation characteristics of the DCN and DCO$^+$ J=3-2 lines are similar, 
the differences in the Fig. 1 images suggest different abundance distributions.
In particular, DCN appears to be more abundant in the inner 
(warmer) disk regions compared to DCO$^+$.

\section{The DCN radial profile}

To constrain the radial distribution of the DCN column density, we use 
the same methods as \citet{Qi08} to model observations
of DCO$^+$, HCO$^+$ and HCN emission from TW Hya.
First, we adopt a physical structure of the TW Hya disk with density 
and temperature distributions from \citet{Qi04,Qi06}.
Second, we assume DCN exists in a disk layer with vertical boundaries 
given by the best-fit values for HCN in \citet{Qi08}; the boundaries of DCN are not very well determined from the present data and chemical models predict a similar vertical distribution of DCN and HCN \citep{Willacy07}.
Third, we model the DCN column density in this layer as a power law, 
$N_{\rm 10}\times(r/10)^{\rm \alpha}$, where  $N_{\rm 10}$ is the column 
density at 10~AU, $r$ the distance from the star in AU, and $\alpha$ the 
power-law index. Based on the DCN data, the model is cut off at 100~AU, which is in agreement with previous models of DCO$^+$ and HCN. For HCN and DCO$^+$, \citet{Qi08} found $\alpha\sim-1$
and $\alpha\sim2$, respectively. 
We therefore tested models of DCN column density with three different
values of power-law index, $\alpha=-1$, 0, and 2, to investigate 
if the DCN distribution can be modeled similarly to DCO$^+$, or if it can
be better described by a flat or decreasing function of radius. 
A more detailed characterization of the shape of the DCN radial profile 
requires higher spatial resolution than currently available.  

\begin{figure}[htp]
\centering
\epsscale{1}
\plotone{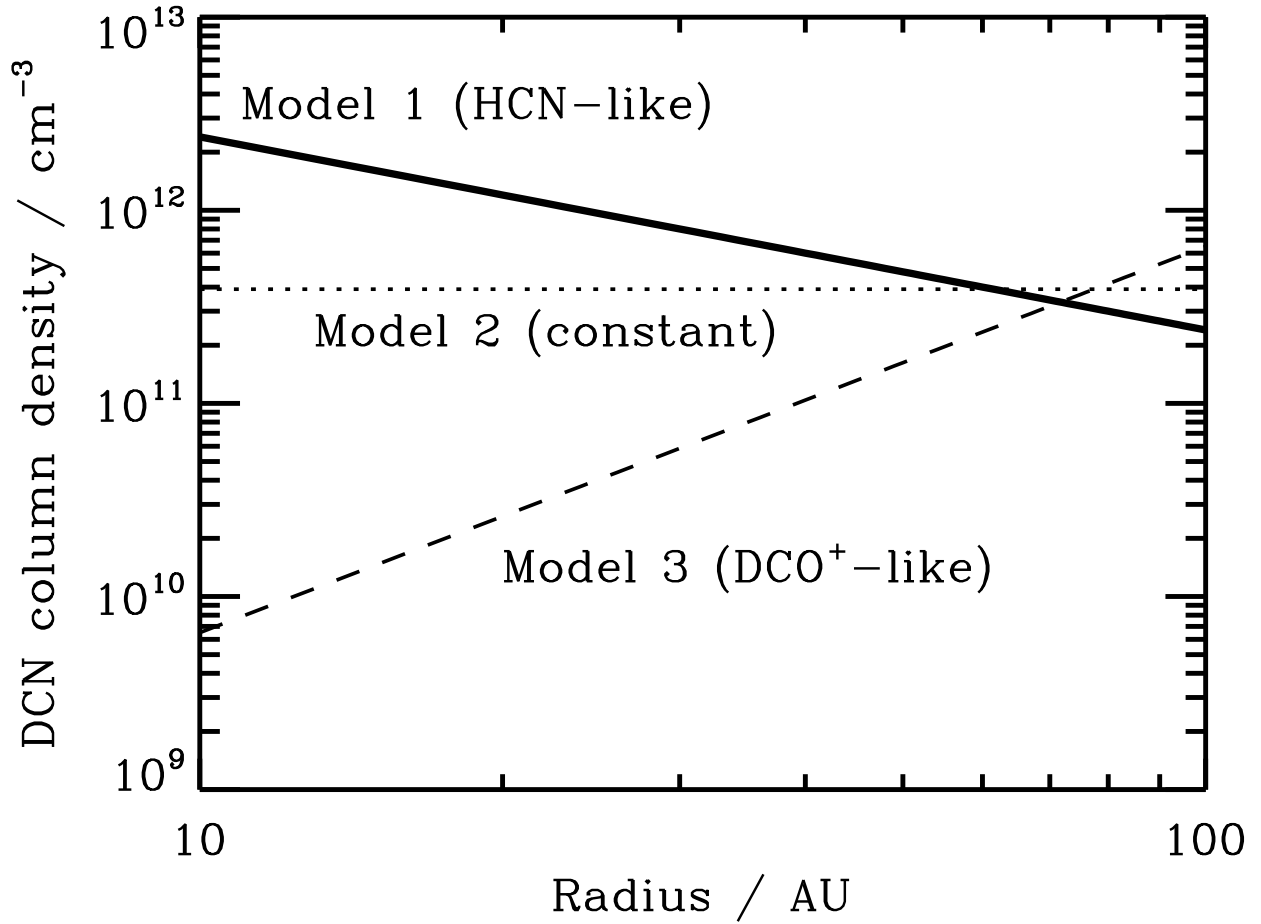}
\caption{
The best-fit radial column density distributions of DCN 
in the disk around TW Hya from 10 to 100~AU for 
a falling power law ($\alpha=-2$, thick line), 
a rising power law ($\alpha=2$ dashed line) and 
a constant ($\alpha=0$ dotted line).
\label{fig2}}
\end{figure}

\begin{figure*}[htp]
\centering
\epsscale{1}
\plotone{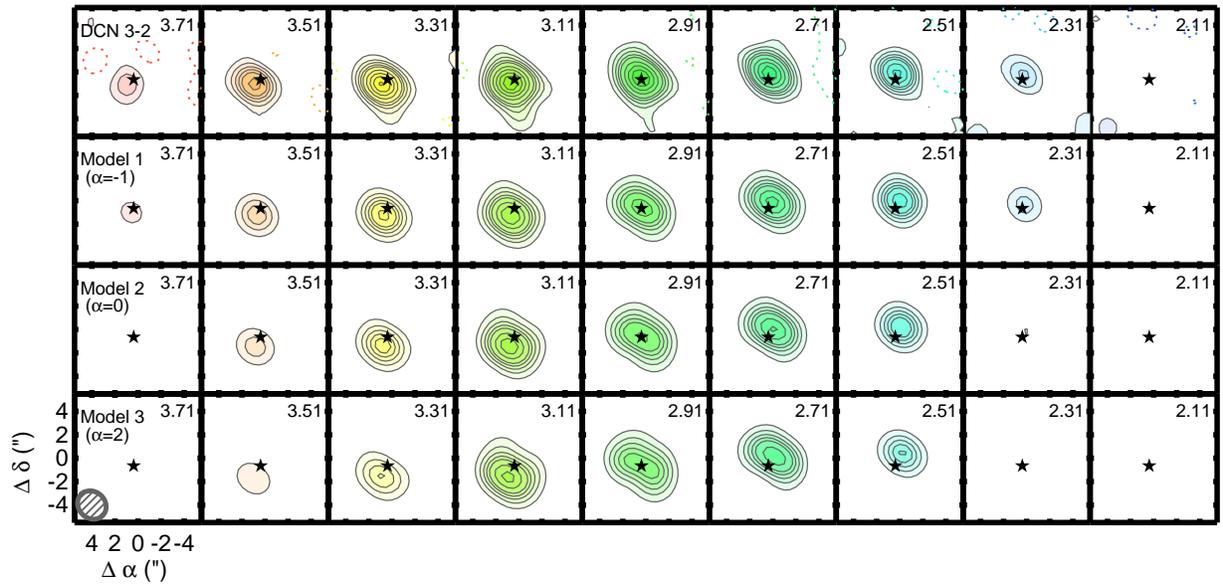}
\caption{
DCN 3-2 velocity channel maps toward TW Hya. The top row shows the observations,
and subsequent rows show models corresponding to power law column density 
distributions with indices $\alpha=-1$ (Model 1), 0 (Model 2) and 2 (Model 3). 
The data is best matched by Model 1, as can be seen by comparing
the shape of the central channels and the extent of extreme velocities. 
The contours are 0.01$\times$[2,4,6,8,10,14,18,22,26,30] Jy beam$^{-1}$.
\label{fig3}}
\end{figure*}

For each power-law model, we optimize $N_{\rm 10}$ by minimizing the
$\chi^2$ value, the weighted differences between the observed and
modeled complex visibilities.  The techniques are described in detail 
in \citet{Qi08}, and the best-fit models shown in Figure~\ref{fig2} 
and Figure~\ref{fig3}. 
Unlike the DCO$^+$, the DCN J=3--2 line is complicated by hyperfine 
structure. The central DCN line quartet at 217.23863~GHz is separated by 0.23~MHz 
(0.32~km~s$^{-1}$) from a singlet at 217.23840~GHz\footnote{http://spec.jpl.nasa.gov/}, with 22\% of the intensity of the 
quartet. This is close enough in frequency and strong enough in relative 
intensity to affect the shape and strength of emission in the
channel maps.  We use the Monte Carlo radiative transfer code RATRAN 
\citep{Hogerheijde00}, taking into account the DCN hyperfine structure, 
to produce model visibilities that are sampled at the appropriate 
spatial frequencies for comparison with the ALMA data. The relative population of the hyperfine levels are assumed to be in LTE, and the relative optical depth of the hyperfine transitions including line overlap is accounted for accurately

Figure \ref{fig3} shows a comparison between the observed DCN J=3-2 
line channel maps and the best fit models for power-law indices 
$\alpha=-1$, 0 and 2. The model with $\alpha=-1$ clearly provides the 
best fit to the data, as models with $\alpha=0$ and $\alpha=2$ underestimate 
the emission in the line wings, show offsets from the observed peaks, and 
appear more elongated in the central channel than the data. Note that the peanut-shaped feature in the systemic velocity channel map of DCN Model 3 is not as obvious as for DCO$^+$ \citep{Qi08} because of the blending of DCN hyperfine components.
The better fit of the model with a negative power index is reflected 
in the $\chi^2$ values of the models, which increase from 2296194 
for Model 1, to 2298334 for Model 2, and to 2300603 for Model 3. 
(Note that we have not excluded the possibility that DCN is even more 
centrally peaked than assumed in Model 1.)

Using Model 1 and the HCN profile reported by \citet{Qi08}, we calculate 
the DCN/HCN abundance ratio to be $\sim0.017$ in the TW Hya disk. 
This result does not depend strongly on the model power-law indices; varying 
the power-law indices by $\pm2$ changes the ratio by less than 30\%. 
This ratio is similar to the previously measured DCO$^+$/HCO$^+$ abundance 
ratio of 0.035 in the TW Hya disk \citep{vanDishoeck03}.  

\section{Discussion}

The observations demonstrate that DCN is centrally peaked on the size
scales observed, ruling out common radial distributions for DCN and DCO$^+$.  
The presence of DCN closer to the star, in regions that are warmer due to 
stellar irradiation, together with the comparable estimates for the 
DCN/HCN and DCO$^+$/HCO$^+$ abundance ratios, suggests a mechanism for 
efficient deuterium enhancement in disk material at higher temperatures 
than implied by reactions solely with H$_2$D$^+$ at T$<$30~K. 
This agrees with standard model predictions, in that DCN can form from gas phase 
reactions involving CH$_2$D$^+$ at T$>$30~K.
However, there are several alternative chemical mechanisms that have the 
potential to affect the relative distributions of DCO$^+$ and DCN, including photochemistry, 
additional gas-phase reactions that result in HCN deuterium fractionation,
grain surface formation of DCN followed by desorption, and molecule specific 
destruction such as depletion onto interstellar grains. Since the origins 
of DCO$^+$ are reasonably well understood, we will focus on the mechanisms 
that affect the DCN distribution.

Before doing so, we note that it is possible for dynamical processes to transport cold 
chemistry products to warmer disk regions. If the destruction rates are slow 
compared to formation, then high deuterium fractionation at $T>30$~K does not 
{\it a priori} imply warm deuterium chemistry. In the context of the DCN and 
DCO$^+$ distributions observed in the TW~Hya disk, both DCN and DCO$^+$ 
could form cold, and then DCO$^+$ could be destroyed during gas diffusion
inwards, while most DCN survives. Such a scenario seems unlikely 
in light of recent models of disk chemistry with diffusion that show 
HCN chemistry is faster than that for HCO$^+$ \citep{Semenov11}. 
A HCN deuterium fractionation mechanism that differs from DCO$^+$ chemistry is thus required to explain the observed DCN distribution.

Photochemistry can produce deuterium fractionation and \citet{Thi10} show that high HDO/H$_2$O ratios in disk gas can be achieved through photochemistry at $T>100$~K, where the HDO formation rate depends on the O+HD reaction. DCN can similarly form through CN+HD. This reactions has a barrier of $\sim$3000~K \citep{Johnston89}, however, and formation of DCN from CN+HD seems unlikely to contribute significantly to the TW Hya DCN distribution on $>50$~AU size scales in the absence of significant vibrational or electronic excitation of the CN. This is consistent with the HDO model results, where the same mechanism seems to be exclusive to the atmosphere of the inner disk where UV fluxes are very high. Since ultraviolet induced photochemistry 
should operate mainly in the disk atmosphere, resolved observations of multiple 
DCN transitions that can constrain the DCN excitation and vertical abundance 
distribution could be used to address the viability of this mechanism.

Freeze-out onto grain surfaces has the potential to regulate the 
DCN distribution relative to DCO$^+$ because of the different binding 
energies of DCN and CO. Gas absorption onto grains is predicted to produce 
molecule-specific ``snowlines'' beyond which abundances drop dramatically. 
This mechanism clearly results in a centrally peaked abundance distribution, 
though warm deuterium chemistry still would be required to produce DCN in 
the first place. It seems unlikely, however, that freeze-out could regulate 
the HCN and DCN distributions on the observed size scales. 
HCN has a comparable adsorption energy to H$_2$O \citep{Aikawa96}, which, 
for the TW Hya disk, results in a snowline at a radius $<10$~AU \citep{Qi08}. 
This does not exclude the possibility that the DCN distribution is directly 
regulated by depletion onto grain surfaces at smaller radii. In fact, 
in the inner hot region of several disks, where HCN is detected at infrared 
wavelengths, the abundances are much higher than observed in outer disks 
at millimeter wavelengths, suggestive of an abundance drop at a few AU due 
to ice formation beyond this radius \citep{Salyk11}. 

Beyond this snowline, ice formation products may be returned to the gas-phase 
through photodesorption \citep[e.g.][]{Oberg09c}. Both laboratory experiments 
and observations of grain surface products show that deuterium fractionation 
can be very efficient in interstellar ices \citep{Nagaoka05,Parise06}. 
Theoretically the DCN/HCN ratio in ices can reach values of $2\times10^{-2}$, 
similar to the ratios predicted from pure gas-phase chemistry 
\citep{Aikawa99,Willacy07}. Recent determinations of low abundances of cold 
H$_2$O gas toward TW Hya and DM Tau \citep{Bergin10,Hogerheijde11} suggest, 
though, that H$_2$O ice evaporation is not efficient outside of a few AU
radius. Because of the low abundances of HCN and DCN relative to H$_2$O, these 
species are likely to act as minor impurities when formed in the ice, and 
desorb only where H$_2$O ice desorbs. Thus, while we cannot rule out ice 
photodesorption as a contributor to the observed DCN distribution, it is 
unlikely to be the dominant process.

Having dismissed dynamics, photochemistry, and mechanisms requiring 
grains, as plausible routes to significant deuterium fractionation in the outer disk, this leaves additional gas-phase formation routes of DCN as the 
main alternative to the reactions involving CH$_2$D$^+$. In the gas-phase, DCN can form 
from HNC+D \citep[e.g.][]{Willacy07}.  The deuterium fractionation in HCN depends then on the gas-phase D/H ratio. The D/H ratio is predicted to be $<10^{-3}$ at large disk radii (250~AU in 
a generic disk model), while it reaches and possibly exceeds 
10$^{-2}$ in the inner disk midplane (17--50~AU) \citep{Willacy07,Willacy09}. 
Thus, reactions between D and HCN provide a plausible pathway to deuterium 
fractionation in HCN in some regions of the disk, and detailed modeling is needed to provide predictions on the vertical and radial abundance profiles resulting from this formation mechanism.

In summary, several reaction and destruction mechanisms may contribute to 
the DCN abundance pattern observed in the TW~Hya disk. 
However, warm CH$_2$D$^+$-driven deuterium chemistry in the inner disk and 
cold H$_2$D$^+$-driven deuterium chemistry in the outer disk provide are
likely to dominate over the alternatives at the disk radii probed by the
observations, and these pathways form consistent picture on their own.
A warm CH$_2$D$^+$-driven deuterium chemistry was also the favored explanation 
for the observations of DCN in the Orion bar \citep{Parise09}.
To determine definitively which formation pathway(s) dominate in disks 
will require interferometric observations with higher spatial resolution, 
observations of related molecules and isotopologs, and more detailed models
of DCN radial distributions for the different chemical scenarios. 
For example, if photodesorption is important, then DCN should spatially 
correlate with other grain surface products, and a powerful photochemistry 
may enhance the $^{15}$N/$^{14}$N ratio in HCN, analogous to what is 
observed in Titan's atmosphere \citep{Liang07}.

Whichever DCN formation pathway dominates, the substantial deuterium 
enhancement in warmer regions of the TW Hya disk challenges the conventional 
wisdom that the high deuterium fractions in comets is necessary evidence for 
a cold (T$<$30~K) chemical history \citep{Mumma11}. 
Most comet deuterium fractions are measured from the D/H ratio in cometary water. In Oort cloud comets  this ratio is $\sim3\times10^{-4}$, enhanced by an order of magnitude compared to the cosmic D/H ratio. Higher D/H ratios of 0.0023$-$0.025 have been observed for DCN/HCN in Hale-Bopp, consistent with models of gas-phase deuteration followed by freeze-out \citep{Meier98,Blake99}. In light of the TW Hya results, even this high ratio does not imply a cold, $T<30$~K origin of the comet material, but may instead reflect the existence a 
lukewarm pathway to deuterium enhancements in the Solar Nebula. The radial distributions
of DCN and DCO$^+$ in disk material around TW Hya and other T Tauri stars 
should be revisited by ALMA at higher resolution to investigate whether the 
observed deuteration chemistry trends persist at smaller scales. Until then, care should be exercised in interpreting the D/H ratios in cometary material.

In general, resolved distributions of molecular emission in disks have 
the potential to put stronger and more direct constraints on key aspects 
of chemical evolution than (global) abundance ratios. In the case of DCN, 
the DCN/HCN abundance ratio of $\sim10^{-2}$ is consistent with a range of 
chemical models that assume different formation pathways.  However, the
observation of different radial distributions of DCN and DCO$^+$ immediately 
suggests the presence of different formation pathways for these species 
and rule out that DCN is mainly formed through reactions with H$_2$D$^+$. 

\acknowledgments

We thank the anonymous referee for detailed comments on especially the disk chemistry, which have improved the paper considerably. The Atacama Large Millimeter/submillimeter Array (ALMA), an international 
astronomy facility, is a partnership of Europe, North America and East Asia 
in cooperation with the Republic of Chile. This paper makes use of the 
following ALMA Science Verification data: ADS/JAO.ALMA\#2011.0.00001.SV. 
Support for KIO is provided by NASA through Hubble Fellowship grant 
awarded by the Space Telescope Science Institute, operated by the Association 
of Universities for Research in Astronomy, Inc., for NASA, under contract 
NAS 5-26555. We acknowledge NASA Origins of Solar Systems grant No. NNX11AK63.

\bibliographystyle{aa}
%\bibliography{mybib}

\end{document}